\documentclass[aps,prd,twocolumn,superscriptaddress,groupedaddress,preprintnumbers,nofootinbib]{revtex4-1}  
\usepackage{graphicx}
\usepackage{bm}
\usepackage{epsf,epsfig}
\usepackage{amscd}
\usepackage{amsmath}
\usepackage{amssymb}
\usepackage{hyperref}
\hypersetup{colorlinks=true,linkcolor=blue,citecolor=red,urlcolor=blue}

\newcommand{\be}{\begin{equation}}
\newcommand{\ee}{\end{equation}}
\newcommand{\bea}{\begin{eqnarray}}
\newcommand{\eea}{\end{eqnarray}}
\newcommand{\hf}{\frac12}
\newcommand{\nn}{\nonumber\\}
\def\eq#1{(\ref{#1})}

\def\hphi{\hat\phi}

\def\vec#1{{\bm{#1}}}
\def\ve{\varepsilon}

\def\CD{\mathcal{D}}
\def\CE{\mathcal{E}}

\def\CG{\mathcal{G}}

\def\CI{\mathcal{I}}
\def\CL{\mathcal{L}}
\def\CM{\mathcal{M}}

\def\CO{\mathcal{O}}
\def\CP{\mathcal{P}}

\def\CT{\mathcal{T}}

\begin{document}

\title{Viscosity and dissipative hydrodynamics from effective field theory}
\author{Sa\v{s}o Grozdanov} 
\affiliation{Rudolf Peierls Centre for Theoretical Physics, University of Oxford, 1 Keble Road, Oxford OX1 3NP, U.K.}
\email{s.grozdanov1@physics.ox.ac.uk}
\author{Janos Polonyi}
\affiliation{Strasbourg University, CNRS-IPHC, 23 rue du Loess, BP28 67037 Strasbourg Cedex 2, France}
\email{polonyi@iphc.cnrs.fr}
\preprint{OUTP-13-10P}


\begin{abstract}
With the goal of deriving dissipative hydrodynamics from an action, we study classical actions for open systems, which follow from the generic structure of effective actions in the Schwinger-Keldysh Closed-Time-Path formalism with two time axes and a doubling of degrees of freedom. The central structural feature of such effective actions is the coupling between degrees of freedom on the two time axes. This reflects the fact that from an effective field theory point of view, dissipation is the loss of energy of the low-energy hydrodynamical degrees of freedom to the integrated-out, UV degrees of freedom of the environment. The dynamics of only the hydrodynamical modes may therefore not posses a conserved stress-energy tensor. After a general discussion of the CTP effective actions, we use the variational principle to derive the energy-momentum balance equation for a dissipative fluid from an effective Goldstone action of the long-range hydrodynamical modes. Despite the absence of conserved energy and momentum, we show that we can construct the first-order dissipative stress-energy tensor and derive the Navier-Stokes equations near hydrodynamical equilibrium. The shear viscosity is shown to vanish in the classical theory under consideration, while the bulk viscosity is determined by the form of the effective action. We also discuss the thermodynamics of the system and analyse the entropy production.
\end{abstract}

\maketitle

\begingroup
\hypersetup{linkcolor=black}
\tableofcontents
\endgroup

\section{Introduction}
Effective theories of gapless long-range (Goldstone) modes are the appropriate framework to systematically derive hydrodynamics \cite{Dubovsky:2005xd,Dubovsky:2011sj,Leutwyler:1996er}. The equations of non-dissipative hydrodynamics have previously been generated using this description at zeroth order in the gradient expansion for relativistic fluids that are insensitive to static, non-compressional deformations \cite{Dubovsky:2005xd,Dubovsky:2011sj} and at second order by \cite{Bhattacharya:2012zx}. This was achieved by constructing a gradient-expanded action describing the long-range scalar modes, which correspond to spatial excitations around the equilibrium state of a fluid. The form of the action was restricted by the identification of appropriate symmetries, with the volume-preserving diffeomorphisms playing the central role in the reduction of potential Lagrangian terms. 

A serious limitation of this scheme is that dissipative forces cannot be derived from the variational principle. Our goal is, however, to develop a systematic scheme for the construction of hydrodynamics at all orders - including dissipation. One approach to this problem is to rely on the linear response theory \cite{Endlich:2012vt}. A different approach aimed at computing hydrodynamic correlation functions from an effective action was recently proposed in \cite{Kovtun:2014hpa,Harder:2015nxa}. In this work, we will report on another method, which will enable us to describe dissipative fluids using the variational principle. This will be done by considering a classical effective action with the characteristics of open system effective field theories, which emerge in the Schwinger-Keldysh Closed-Time-Path (CTP) formalism \cite{schw,Keldysh:1964ud}, first introduced by Schwinger \cite{schw}. The formalism was invented to describe retarded time-evolution of operator expectation values acting on mixed states, which are specified by density matrices. 

We will begin this paper by presenting the CTP formalism as an extension of the usual quantum field theory used to compute scattering amplitudes between asymptotic pure states. We will focus on the matrix structure of the CTP propagators, which arises from the doubling of the degrees of freedom and the introduction of {\it two time axes}; one evolving from past to future and the other evolving backwards in time. Effective theories emerge when the unobserved degrees of freedom, called the {\it environment}, are eliminated. The remaining degrees of freedom, called the {\it system}, follow more involved effective dynamics than in a theory of pure states. This requires the use of the CTP formalism, which is able to incorporate interactions and entanglement between the system and the environment. As a result of its original interactions with the environment, the dynamics of the system cannot conserve energy. This view is consistent with the effective field theory understanding of {\it dissipation}, as the energy loss of the IR macroscopic degrees of freedom to the integrated-out, UV microscopic degrees of freedom of the environment.  

We will argue that, generically, CTP effective field theories include couplings between the two time axes, expressed within the {\it influence functional} considered first by Feynman and Vernon in \cite{Feynman:1963:TGQ}, which includes all effective interactions. The coupling of the two time axes is a result of the system-environment interactions, after the environment had been integrated out, with the details of the effective action depending both on the choice of the integrated-out degrees of freedom as well as the initial state. This in turn leads to an effective theory of the low-energy system, which experiences dissipative dynamics.

This structure descends into a classical low-energy theory, which we will use to derive dissipative hydrodynamical equations of motion from variational principle. This will be done at a phenomenological level, directly in terms of an effective classical CTP field theory without a microscopic derivation, in accordance with the logic used in \cite{Dubovsky:2005xd,Dubovsky:2011sj}. By varying the fields on only one of the two CTP time axes, we will obtain the energy-momentum balance equation containing a two-tensor that will not be conserved because of interactions between the IR degrees of freedom of the fluid and the environment. Near hydrodynamical equilibrium, however, we will show that this tensor becomes approximately conserved. We will then discuss the dynamical regime in which this tensor can be identified with the fluid's phenomenological stress-energy tensor. Using the energy-momentum balance equation, we will derive the equations of motion, which will have the form of the Navier-Stokes equations. The shear viscosity will be shown to vanish and a possible cause of this restriction will be discussed, i.e. the theory's invariance under volume-preserving diffeomorphisms. Non-vanishing thermodynamical quantities and the bulk viscosity will be identified in terms of the coefficient functions of the effective Lagrangian.   

In the phenomenological approach based on the CTP formalism, which we will employ in this paper, the equations of motion and the energy-momentum balance equation are closed without thermodynamical considerations. This means that thermodynamical variables can only be identified by using the algebraic structure of the energy-momentum balance equation or the full stress-energy tensor in the non-dissipative case. In a self-consistent theory of hydrodynamics, such considerations should automatically lead to positive entropy production. What we shall see is that this is not automatically ensured by the effective action analysed in this work and that constraints have to be imposed on its form even near the hydrodynamical equilibrium. We will comment on how such considerations can be avoided in effective theories derived from a unitary microscopic quantum field theory. Our analysis of entropy will be performed only near equilibrium, where the stress-energy tensor is approximately conserved. We will defer a more complete discussion of entropy in the framework of the CTP effective actions to future work.

Finally, we will conclude by summarising our results.

\section{CTP formalism and effective QFT}\label{sec:CTPinQFT}

The CTP formalism was initially introduced to facilitate a computation of an expectation value $\langle A, t_1 | \CO (t_2) | A,t_1 \rangle$ at time $t_2$, given an initial (mixed) state $A$, or a density matrix $\rho_i$, at time $t_1$ \cite{schw}. By inserting two complete sets of states,
\begin{align}\label{ExpO}
&\langle A, t_1 | \CO (t_2) | A,t_1 \rangle =  \nn
&= \sum_{B,B'} \langle A, t_1 | B,t_2 \rangle\langle B,t_2| \CO (t_2) | B',t_2 \rangle\langle B',t_2 | A,t_1 \rangle,
\end{align}
expression \eqref{ExpO} can be interpreted as an evolution of $A$ from $t_1$ to $t_2$, when the trace is evaluated, accompanied by a time evolution from $t_2$ backwards in time to $A$ at $t_1$. The two time axes thus both include information about the physical system, which leads to the doubling of degrees of freedom. We introduce notation 
\begin{align}
\varphi \to \hat\varphi = (\varphi^+,\varphi^-),
\end{align}
with $\varphi^+$ and $\varphi^-$ propagating on separate time axes. The fields $\varphi^\pm$ are identified at some final time $t_f > t_2$, i.e. $\varphi^+(t_f) = \varphi^- (t_f)$. 

Let us consider a scalar field $\varphi$ with the single time axis action $S_s [\varphi]$. The doubling of the degrees of freedom leads to the quantum generator functional,
\begin{align}\label{CTPGenFunc}
e^{i W_{CTP}[\hat J]} = \int  \CD\hat\varphi \exp\left\{ i S_s[\varphi^+] - i S_s [\varphi^-] + i \int  \hat J \hat\varphi   \right\}.
\end{align}
The full CTP action of $\hat\varphi$, 
\begin{align}\label{CTPActQFT}
S_{CTP} [\varphi^+,\varphi^-] = S_s[\varphi^+] - S^*_s [\varphi^-], 
\end{align}
possesses the CTP symmetry,
\begin{align}\label{CTPSymmQFT}
S_{CTP}[\varphi^+,\varphi^-]=-S^*_{CTP}[\varphi^-,\varphi^+].
\end{align}
The generator functional with two sources $\hat J = (J^+,J^-)$, leads to a $2\times2$ matrix propagator,
\begin{align}\label{bosctppr}
i\hat D(x,y)=\begin{pmatrix}\left\langle \CT\left[\varphi(x)\varphi(y)\right]\right\rangle &\left\langle \varphi(y)\varphi(x)\right\rangle \cr\left\langle\varphi(x)\varphi(y)\right\rangle&\left\langle \CT \left[\varphi(y)\varphi(x)\right]\right\rangle^*\end{pmatrix},
\end{align}
where $\CT$ denotes the time ordering in the Feynman propagator and $\CT^*$ the anti-time ordering. The bosonic propagator can be further written in the form of
\begin{align}\label{stctpform}
\hat D=\begin{pmatrix}D^n+iD^i&-D^f+iD^i\cr D^f+iD^i&-D^n+iD^i\end{pmatrix},
\end{align}
where the near, $D^n$, and far, $D^f$, Green's functions give the retarded and advanced propagators, $D^{ret} =D^n + D^f$ and $D^{adv} = D^n - D^f$. The free propagator contains the Feynman propagator as the diagonal block, $D^{++}$, and the Wightman function, 
\begin{align}
\left\langle \varphi(y)\varphi(x)\right\rangle =  - 2\pi i\delta(k^2 + m^2)\Theta(-k^0).
\end{align}
The off-diagonal entries of the propagator $\hat D$ induce interactions between $\varphi^+$ and $\varphi^-$. Finite temperature, $T = 1/\beta$, and density with a chemical potential $\mu$, in cases when $\varphi$ is complex, further modify the free propagator by a relation, $\hat D_0\to \hat D_0 + \hat D_{T,\mu}$, where
\begin{align}
i \hat D_{T,\mu}(k) = -2\pi i \delta\left(k^2 + m^2\right)n_B(k)\begin{pmatrix}1&1\cr1&1\end{pmatrix},
\end{align}
and the Bose-Einstein distribution is 
\begin{align}
n_B(k) =\frac{\Theta(-k^0)}{e^{\beta(\epsilon_\vec{k}+\mu)}-1} +\frac{\Theta(k^0)}{e^{\beta(\epsilon_\vec{k}-\mu)}-1} . 
\end{align}

Let us now consider a massive scalar field theory with a $\lambda \varphi^4$ coupling, where $\lambda$ is treated as a small perturbative coupling constant. We wish to follow the Wilsonian approach to effective field theory and integrate out the UV-degrees of freedom.\footnote{In this paper, we only consider the schematic structure of effective actions and leave the details of the effective Wilsonian $\varphi^4$ action for future work. We note that the vast enhanced complexity of the CTP effective actions can be seen from the fact that a simple $\varphi^4$ action will include eight real couplings, $\Re m$, $\Im m$, $\Re \lambda$, $\Im \lambda$, $\Im \mu_1$, $\Im\mu_2$, $\Re\mu_3$ and $\Im\mu_3$, up to quartic order. } We introduce a scheme with two cut-offs in the original bare theory, one for frequency, $| k_0 | < \Lambda_0$, and one for momentum, $\sqrt{\vec{k}^2+ m^2} = \ve_{\vec{k}} < \Lambda_\ve$. We then split the fields $\hat\varphi = \hat\varphi_<+ \hat\varphi_>$ and integrate out $\hat\varphi_>$, with frequency {\it or} energy in the following regions, 
\begin{align}
\xi \Lambda_0 \leq | k_0 | < \Lambda_0, && \zeta \Lambda_\ve \leq \ve_{\vec{k}} < \Lambda_\ve.
\end{align}
The UV-mode integrals must thus run over three regions,
\begin{align}
I_1:&~~~~\left\{\xi \Lambda_0 \leq k_0  < \Lambda_0 ,\, 0  \leq \ve_{\vec{k}} < \Lambda_\ve\right\}, \\
I_2:&~~~~\left\{-\Lambda_0 < k_0  \leq - \xi \Lambda_0 ,\, 0  \leq \ve_{\vec{k}} < \Lambda_\ve\right\}, \\
I_3:&~~~~\left\{- \Lambda_0 < k_0  < \Lambda_0 ,\,  \zeta\Lambda_\ve  \leq \ve_{\vec{k}} < \Lambda_\ve\right\}.
\end{align}
In a perturbative expansion of \eqref{CTPGenFunc}, we find various couplings between the two axes, for example $\lambda^2 \left( \varphi^+_< \right)^2 \left(\varphi^+_>\right)^2  \left( \varphi^-_< \right)^2 \left(\varphi^-_>\right)^2$. In the process of integrating out $\varphi_>^+$ and $\varphi_>^-$, the on-shell Wightman functions connect vertices on different time axes, and give rise to non-trivial $\varphi_<^+ \varphi_<^-$ couplings in the effective theory, $S_{eff} [\hat\phi_<]$. We find that the effective action includes the following type of terms, 
\begin{align}
&S_{eff} [\hat\phi_<] = S_{CTP} [\phi^+_<, \phi^-_<] + \int d^4 x \left[\mu_1 \, \varphi^{+}_<  \varphi^{-}_< \right.\nn
&\left.+  \mu_2 \, \varphi^{+2}_<  \varphi^{-2}_<  + \mu_3  \, \varphi^{+3}_< \varphi^-_> - \mu_3^* \, \varphi^{-3}_< \varphi^+_>  + \ldots \right]. 
\end{align}
Due to the CTP symmetry, $\mu_1$ and $\mu_2$ have to be purely imaginary, whereas $\mu_3$ will be complex. The equations of motion for $\hat\phi$ derived from a CTP effective action will thus also in general be complex. The real part of the equations of motion, coming from $\Re S_{eff}$, has the property that $\phi^+ = \phi^-$ is the solution, which is always true in real CTP actions. The imaginary terms from $\Im S_{eff}$ will be the complex conjugates of each other in the equations for $\phi^+$ and $\phi^-$. We should note that the same structure as in the Wilsonian effective action would also arise in a 1PI effective action. In both effective actions the real part of the action is important for physical Hermitian expectation values, whereas the imaginary part controls decoherence. 

Beyond this proof of principle, which shows that coupling between $\phi^+$ and $\phi^-$ generically arise in effective actions, we will discus the significance of such effective coupling in the following section. Furthermore, note that this type of effective theory, which is constructed with the full CTP machinery, is able to account for the time-evolution of any pure or mixed state in a closed or open field theory system. The microscopic generator functional with a non-trivial initial density matrix would be written as
\begin{align}\label{GenFunInState}
e^{i W_{CTP}[\hat J]} =&  \! \int \! \CD \hat\varphi \, \rho_i \left[\hat\varphi(t_i,\vec{x}) \right]   \exp\left\{ i S_{CTP}[\hat\varphi] + i \int  \hat J \hat\varphi   \right\},
\end{align}
instead of Eq. \eqref{CTPGenFunc}. The details of the effective system we are describing are thus determined by the degrees of freedom that were integrated out, i.e the {\it environment}, as well as the choice of the initial state. The remaining {\it reduced density matrix} of the sub-{\it system} encodes all of the information about the entanglement with the environment and dissipation of energy from the sub-system. The sub-system can thus either preserve or break various symmetries of the full closed system. Of particular relevance to us will be the fact that energy and momentum are no longer conserved in an effective theory of such a system. More precisely, one can no longer find a conserved Noether current, which corresponded to translational invariance in the full microscopic theory.

\section{CTP formalism in classical field theory}

\subsection{Closed system}\label{sec:ClassicalClosedSystem}
To see how the features presented in Section \ref{sec:CTPinQFT} can also follow directly from a classical theory, let us consider a classical field theory for an isolated system, described by the field $\psi(x)$, which is invariant under time inversion. Instead of deriving the effective theory for an open system from microscopic dynamics, we can use the CTP formalism in classical physics \cite{clctp,Galley:2012hx}. This is necessary when considering a physical problem in which we wish to specify the initial conditions for the equations of motion and to have the possibility of introducing effective interactions with dissipative forces into the Lagrangian formalism. From the microscopic point of view, such a theory can be understood as an effective field theory, a special case of those considered in Section \ref{sec:CTPinQFT}, which keep the IR dynamics closed. All of the consideration below would follow directly form such a derivation.

The procedure again begins by doubling the degrees of freedom \cite{bateman}, 
\begin{align}
\psi\to\hat\psi=(\psi^+,\psi^-), 
\end{align}
in a way that both members of the CTP doublet satisfy the same equation of motion, initial conditions and the relation $\psi^+(t_f,\bm{x})=\psi^-(t_f,\bm{x})$ at the final time. The action describing the dynamics of $\hat\psi$ is defined as in \eqref{CTPActQFT},
\begin{align}\label{ctpaction}
S_{CTP}[\hat\psi] =  \int_{t_i}^{t_f}  d^{d+1} x \left\{ \CL_s \left[\psi^+\right] - \CL^*_s \left[\psi^- \right] \right\},
\end{align}
where $\CL_s\left[\psi\right] = \CL \left[ \psi,\partial\psi \right] +i\epsilon\psi^2$ differs from the original Lagrangian in that it splits the degeneracy of the CTP action for $\psi^+(x)=\psi^-(x)$. In the expression for $\CL_s$, $\epsilon \ll 1$ is an infinitesimally small number. The action \eq{ctpaction} possesses the CTP symmetry \eqref{CTPSymmQFT} related to the exchange of the two time axes, $\psi^+\leftrightarrow\psi^-$, and implies the relation 
\begin{align}\label{CTPSymmetryMic}
S_{CTP}[\psi^+,\psi^-]=-S^*_{CTP}[\psi^-,\psi^+],
\end{align}
which must be obeyed by any classical CTP action. 

\subsection{ Open systems }

In order to describe an {\it open system} of the IR gapless hydrodynamical degrees of freedom in the language of classical field theory, we first need to consider a question of how to construct a general classical field theory of a subset $\phi$ of the degrees of freedom $\psi$. The effective dynamics of $\phi$ can be obtained by eliminating the environment degrees of freedom by using their equations of motion. Similarly, from the point of view of QFT presented in Section \ref{sec:CTPinQFT}, the environment could be seen as the degrees of freedom, which are integrated out. This view is consistent with the microscopic view of dissipation in hydrodynamics; it is the energy loss of the fluid's IR degrees of freedom coupled to the UV degrees of freedom of the environment. Only the total closed system, combining all degrees of freedom, conserves energy. 

In classical CTP theory, one finds that the same structure arises as in Section \ref{sec:CTPinQFT}. The effective action, which results from this procedure, again has a more involved structure than \eq{ctpaction}, namely
\begin{align}\label{effaction}
S_{eff} [\hphi ]=S_1 [\phi^+] - S_1^* [\phi^- ]+S_2 [\hphi],
\end{align}
where the indices $1$ and $2$ reflect the number of time axes entering the term in the action. $S_1$ and $S_2$ can be uniquely distinguished by imposing
\begin{align}
\frac{\delta^2 S_2}{\delta\phi^+\delta\phi^-} \ne 0.
\end{align} 
Elimination of the environment generates contributions both to $S_1$ and $S_2$. We would like to point out that in the original terminology of Feynman and Vernon \cite{Feynman:1963:TGQ}, {\it all} effective contributions to $S_{eff}$ were collected into the influence functional $S_i$,
\begin{align}\label{FVinf}
S_{eff} = S_0[\phi^+] - S_0^*[\phi^-] + S_i [\hat\phi].
\end{align}
In Eq. \eqref{FVinf}, $S_0$ stands for the original single time-axis action preceding the elimination of the environment. We find it is more convenient to separate the influence functional into terms entering $S_1$ and $S_2$. In this language, $S_0$ will be included in $S_1$. This separation is useful because the terms in $S_1$ preserve energy and momentum, while terms in $S_2$ represent dissipative forces. The inclusion of $S_2$ into the classical action for hydrodynamics, discussed in Section \ref{sec:HydroDissipation}, will thus be our addition to the previous works on deriving hydrodynamics from an action principle \cite{Dubovsky:2005xd,Dubovsky:2011sj,Bhattacharya:2012zx}. 

In the classical picture, the couplings between $\phi^+$ and $\phi^-$ appear due to the boundary conditions for the environment coordinates at the final time. These contributions arise from asymptotic long-time excitations of the environment and are usually approximated by gradient expansion. The imaginary part of the effective action obtained by eliminating the environment remains infinitesimal, as in the case of an isolated system. It will be ignored below.

Let us assume that the gradient expansion in terms of space-time derivatives is applicable in the effective action \eqref{effaction}. We impose identical initial conditions on the two time axes, $\partial^n_{t}\phi^+(t_i,\bm{x})=\partial^n_{t}\phi^-(t_i,\bm{x})$, together with the auxiliary conditions $\partial^n_{t}\phi^+(t_f,\bm{x})=\partial^n_{t}\phi^-(t_f,\bm{x})$ for all orders of derivatives labeled by $n\geq 0$. 

Variational equations can thus still be derived in the CTP theory because the boundary contributions arising from partial integration cancel, due to the above conditions. Furthermore, the solutions of the open system's Euler-Lagrange equations of motion give 
\begin{align}
\phi^+(x)=\phi^-(x).
\end{align} 
The CTP symmetry \eqref{CTPSymmetryMic} implies that any effective action must also obey the same symmetry,
\begin{align}\label{CTPSymmetryMic}
S_{eff}[\phi^+,\phi^-]=-S^*_{eff}[\phi^-,\phi^+].
\end{align}
From the point of view of effective field theory, relation \eqref{CTPSymmetryMic} can be seen as a constraint on the form of terms one can write down in the effective action.

As an example of this formalism, it is instructive to consider a non-relativistic one-dimensional particle whose effective theory is defined by the Lagrangians
\begin{align}
&\CL_1=\hf \left(m\dot x^2 - m\omega^2x^2\right), \\
&\CL_2=\frac\gamma2 \left(x^-\dot x^+  - x^+\dot x^-\right),
\end{align}
where $\gamma$ is an arbitrary coupling constant. The corresponding equations of motion describe a damped harmonic oscillator,
\begin{align}
m\ddot x^\pm+\gamma\dot x^\mp+m\omega^2x^\pm = 0,
\end{align}
hence
\begin{align}
x \equiv x^+ = x^-.
\end{align}
The coupling constant $\gamma$ in the influence functional is thus controlling the friction term inside the on-shell equation of motion, 
\begin{align}
m\ddot x +\gamma\dot x +m\omega^2x  = 0. 
\end{align}
It is clear that the conservation of energy is violated by $\CL_2$.

In CTP, the na\"{i}ve application of the Noether theorem to the action \eq{effaction} gives, due to the CTP symmetry, an identically vanishing stress-energy tensor for fields that satisfy the equations of motion. However, the trivial cancellation between the time axes can be avoided and the balance equation can be derived by varying only one of the CTP doublet fields,
\begin{align}
\label{ctpvar}
\begin{split}
&\phi^+(x)\to\phi^+(x+a(x)), \\
&\phi^-(x)\to\phi^-(x).
\end{split}
\end{align}
The equation of motion for $a(x)$, {\em the balance equation}, can then be written in the form of a tensor divergence as
\begin{align}\label{embalance}
\partial_\mu T^{\mu\nu}=R^\nu.
\end{align}
Note that the dynamics of $\phi^+$ and dynamics of the $\phi^-$ degrees of freedom on the two time axes are related to each other by the CTP symmetry, \eqref{CTPSymmetryMic}. Either time axis could thus be used in this construction. In this work, we will always choose to treat the positive axis with $\phi^+$ fields as the one directly relevant to physical observations.  

\section{Hydrodynamics}
An effective field theory describing hydrodynamics has recently been developed in terms of a gradient expansion of the gapless IR Goldstone modes arising from the broken spatial boost invariance \cite{Dubovsky:2005xd,Dubovsky:2011sj}. Reference \cite{Nicolis:2013lma} used the coset construction of a space-time symmetry breaking pattern to show that three scalar modes were sufficient in parametrising the low energy effective theory. The dynamics of the scalar modes $\phi^I$, with $I=\{1,2,3\}$, in flat $3+1$ dimensional space-time with metric $\eta_{\mu\nu}=\text{diag}\left(-1,1,1,1\right)$ displays internal symmetries under rigid translations, 
\begin{align}
\phi^I\to\phi^I+\alpha^I, ~~ \text{with}~ \alpha^I=const.,
\end{align} 
rotations, 
\begin{align}
\phi^I\to R^I_J \phi^J, ~~\text{with}~ R^I_J \in SO(3),
\end{align} 
and volume-preserving diffeomorphisms (reparametrisations), which we abbreviate by $S\text{Diff}\,(\mathbb{R}^{1,3})$,
\begin{align}
\phi^I\to\xi^I(\phi), ~~\text{with}~ \det \left( \frac{ \partial\xi^I }{ \partial \phi^J }\right) = 1. 
\end{align}

The $S\text{Diff}$ symmetry, which is imposed here, deserves special attention. Arnold showed that non-dissipative ideal hydrodynamical equation on a manifold $\CM$, i.e. the Euler equation, can be generated as the co-adjoint orbit on the Lie group manifold of $S\text{Diff}(\CM)$ \cite{Arnold1,Arnold2}. This symmetry should be broken by dissipation, but this mechanism has not been understood. We will proceed by making use of it and comment at the end on why this symmetry is most likely too restrictive to construct the full equations of viscous fluids.

Returning to the setup of \cite{Dubovsky:2005xd,Dubovsky:2011sj}, we note that in equilibrium, the fields equal the spatial coordinates $\phi^I = const. \cdot x^I$. Furthermore, relativistic hydrodynamics also requires the Poincar\'{e} symmetry. The gradient expansion is constructed by counting the number of derivatives acting on the vector field,
\begin{align}\label{K}
K^\mu  &=  \frac{1}{6} \epsilon^{\mu\alpha_1\alpha_2\alpha_3} \epsilon_{IJK} \partial_{\alpha_1} \phi^I\partial_{\alpha_2} \phi^J \partial_{\alpha_3} \phi^K  \nn
&\equiv  P^{\mu\alpha}_K \partial_\alpha \phi^K, 
\end{align}
which is a combination of gradients of the Goldstone modes allowed by the symmetries in three spatial dimensions. The vector field is conserved because of its anti-symmetric structure, 
\begin{align}\label{Kconser}
\partial_\mu K^\mu=0,
\end{align}
and keeps the comoving coordinates constant along its direction, $K^\mu\partial_\mu\phi^I\! = 0$. We can introduce a scalar field $b$, such that 
\begin{align}
K^\mu\equiv bu^\mu.
\end{align} 
The norm $u^\mu u_\mu = -1$ then implies that $b^2 = - K^\mu K_\mu$. 

Two useful projector identities can be derived for $P^{\mu\alpha}_K$ as defined in \eqref{K} by using the properties of $K^\mu$,
\begin{align}
&P^{\mu\nu}_K \partial^\lambda \phi^K = \frac{1}{3} \left( K^\mu \Delta^{\nu\lambda} - K^\nu \Delta^{\mu\lambda} \right) , \label{ProjId} \\
&P^{\mu\nu}_K \partial^\lambda \partial_\nu \phi^K = \frac{1}{3} \partial^\lambda K^\mu , \label{ProjId2}
\end{align}
with $\Delta^{\mu\nu}=\eta^{\mu\nu} + u^\mu u^\nu$. The zeroth and first-order Lagrangians for the uncharged fluid are then
\begin{align}
\CL^{(0)} + \CL^{(1)}  = F(b) + g(b) K^\mu K^\nu \partial_\mu K_\nu. \label{Lag01}
\end{align}
At zeroth order \cite{Dubovsky:2011sj}, the conserved stress-energy tensor of the closed system takes the form of an ideal fluid,
\begin{align}
T_{(0)}^{\mu\nu}=\varepsilon u^\mu u^\nu+p \Delta^{\mu\nu}, \label{TnonCTP}
\end{align}
where the energy density $\varepsilon(x)$ and pressure $p(x)$ are space-time dependent functions, which are directly determined by the function $F(b)$ in the Lagrangian by
\begin{align}
&\varepsilon=-F , \\
&p=F-b\partial_bF.
\end{align} 
Further thermodynamic analysis reveals that the temperature is given by 
\begin{align}\label{temp}
T=-\partial_bF.
\end{align} 
Finally, the vector field $K^\mu$ was interpreted at this order as the conserved entropy current of ideal hydrodynamics \cite{Dubovsky:2011sj},
\begin{align}\label{EntropyCurrentK}
K^\mu = b u^\mu \equiv S^\mu = s u^\mu,
\end{align} 
with 
\begin{align}
s = b
\end{align} 
being the entropy density. This identification was performed in \cite{Dubovsky:2011sj} because $K^\mu$ is parallel to $u^\mu$ and is by construction conserved, which is consistent with the entropy conservation in an ideal fluid, i.e. in zeroth-order relativistic hydrodynamics. The above results are also consistent with the usual thermodynamical relations, $\varepsilon + p = s T$, $T = \partial \varepsilon / \partial s$ and $s = \partial p / \partial T$.

In reference \cite{Bhattacharya:2012zx}, the authors considered non-dissipative second-order hydrodynamics by using the same identification of the entropy current, noting that the construction should be understood as being done in the {\it entropy frame}, in which $S^\mu = s u^\mu$ to all orders in the absence of dissipation. In standard phenomenological hydrodynamics, one instead of the entropy frame usually chooses either the Landau frame or the Eckart frame \cite{Kovtun:2012rj}. The physical meaning of the Landau frame is that there is no energy flow in the local rest frame of the fluid. The Eckart frame, useful for a description of charged fluids, means that there is no charge flow in the local rest frame. 

The first-order contribution to the Lagrangian \eqref{Lag01} can be rewritten as a total derivative and hence does not contribute to $T^{\mu\nu}$. As a final point in this construction, note that the chemical potential is vanishing in the absence of a conserved $U(1)$ Noether current \cite{Dubovsky:2011sj}, which we will not consider in this work.

\section{Hydrodynamics with dissipation}\label{sec:HydroDissipation}
\subsection{The setup}

Variational methods in the usual effective theory formalism cannot describe dissipation. However, this limitation can be avoided by using the CTP scheme as introduced above. Firstly, the degrees of freedom are doubled, giving us six Goldstone fields $\phi^{\pm I}$. The action must be invariant under pairs of translations, rotations and volume-preserving diffeomorphisms, each acting independently on $\phi^{I+}$ and $\phi^{I-}$. The diffeomorphisms act as 
\begin{align}
\phi^{\pm I}  \to \xi^{\pm I} \left( \phi^\pm \right), 
\end{align}
with two independent conditions on the determinants,
\begin{align} 
\det \left( \frac{\partial\xi^{+ I} }{ \partial \phi^{+ J} } \right) = 1, && \det \left( \frac{ \partial\xi^{- I} }{ \partial \phi^{- J}} \right) =
 1. 
\end{align}
 The field content and symmetries allow for two independent currents $K^{i\mu}$, both with the same Lorentz structure as before, where $\{i,j,k,...\}\in\{0,3\}$ correspond to the number of $\phi^+$ fields inside $K^{i\mu}$,
\begin{align}\label{KCTP}
K^{i\mu} &= \frac{1}{6} \epsilon^{\mu\alpha_1\alpha_2\alpha_3} \epsilon_{IJK} \partial_{\alpha_1} \phi^{\sigma_1I} \partial_{\alpha_2} \phi^{\sigma_2 J} \partial_{\alpha_3} \phi^{\sigma_3 K},
\end{align}
with $(\sigma_1\sigma_2\sigma_3)=\{(---),(+++)\}$ for $i=\{0,3\}$. Both $K^{i\mu}$ are still conserved, 
\begin{align}\label{KconservCTP}
\partial_\mu K^{i\mu} = 0, 
\end{align}
and both $K^{i\mu} = K^\mu$ after $\phi^{+K} = \phi^{-K}$ is imposed. It is useful to define, as in Eq. \eqref{K}, 
\begin{align}\label{P3def}
K^{3\mu} \equiv P^{3\mu\alpha}_K \partial_\alpha \phi^{+K}.
\end{align}
Furthermore, we can introduce 
\begin{align}\label{P0def}
P^{0\mu\alpha}_K \equiv 0,
\end{align}
which will make it clear that the transformation $\delta^+$ acting on $K^{0\mu}$ gives a vanishing contribution.

We can now write down the CTP action for the first two orders in the gradient expansion of $K^{i\mu}$, 
\begin{align}
&\CL^{(0)}_{CTP}=F(K^3_\gamma K^{3\gamma})-F(K^0_\gamma K^{0\gamma})+G(K^i_\gamma K^{j\gamma}),   \label{LagCTP1}  \\
&\CL^{(1)}_{CTP}=\sum_{i,j,k} f_{ijk}(K^l_\gamma K^{m\gamma})K^{i\mu}K^{j\nu}\partial_\mu K^k_\nu.\label{LagCTP2}
\end{align}
Small latin indices are always summed over $\{0,3\}$. The single axis contributions, i.e. $S_1$, to $\CL^{(0)}$ remain the same as in \eqref{Lag01} and the zeroth-order action $S_2$, which includes couplings between the two time axes, is parametrised by $G$. It mixes $K^{i\mu}$'s with different CTP indices. We include no single axis action at first order, as it would be a total derivative \cite{Dubovsky:2011sj}, so $\CL^{(1)}$ is purely a part of $S_2$, as classified by Eq. \eqref{effaction}. This means that $f_{333}$ cannot be a function of only $K^{3\mu}$ and $f_{000}$ not of only $K^{0\mu}$. The real coefficient functions $F$, $G$ and $f_{ijk}$ can depend on any Lorentz-contracted combination of $K^{i\mu}$, but may include no derivatives. At first order, we thus have $2^3=8$ coefficient functions $f_{ijk}$, which are reduced to $4$ independent functions by the CTP symmetry \eqref{CTPSymmetryMic}.

\subsection{Energy-momentum balance equation }

The variation of the current $K^{i\mu}$ with respect to $\phi^+$ results in an expression that is weighted by the number of $\phi^+$ fields inside of $K^{i\mu}$,
\begin{align}
\delta^+_\phi K^{i\mu} = i P^{i \mu \alpha}_K \partial_\alpha \delta \phi^{+K}. 
\end{align}
The zeroth-order Euler-Lagrange equations of motion are
\begin{align}\label{EOM0}
\partial_\lambda \sum_{i\leq j} \frac{\partial\left(F+G\right)}{\partial ( K^i_\alpha K^{j\alpha} )} \left(i P^{i \mu\lambda}_K K^j_\mu  + j K^i_\mu P^{j\mu\lambda}_K \right)  = 0.
\end{align}
To find the energy-momentum balance equation for the open system, we vary the space-time dependence of $\phi^+$ by $x \to x + a(x)$. This results in $\delta^+_x \phi^{+K} = a^\mu \partial_\mu \phi^{+K}$, while leaving $\delta^+_x \phi^{-K} = 0$. By using the definitions of $K^{i\mu}$ as stated in Eqs. \eqref{KCTP}, \eqref{P3def} and \eqref{P0def}, it follows that
\begin{align}
\delta^+_x K^{i\mu} = i P^{i\mu\alpha}_K \left(\partial_\alpha a_\lambda \partial^\lambda \phi^{+K} + a_\lambda \partial^\lambda \partial_\alpha \phi^{+K} \right).
\end{align}
After we identify $\phi^{+K} = \phi^{-K}$, which is implied by the equations of motion, and use projector identities \eqref{ProjId} and \eqref{ProjId2}, the form of the left-hand-side of \eqref{embalance} remains that of $T_{(0)}^{\mu\nu}$ in \eqref{TnonCTP}. The energy density and pressure are now
\begin{align}
&\varepsilon = - F, \label{CTP0coeff1}  \\
&p = F - b\partial_b F + \frac{b^2}{3} \sum_{i\leq j } \bar G'_{ij} \left(i+j\right),\label{CTP0coeff2}
\end{align}
and the non-conserved part of the balance equation is 
\begin{align}\label{force0}
R_{(0)}^{\nu} = \sum_{i\leq j} \bar G'_{ij} \left(i+j\right)   b \partial^\nu b / 3. 
\end{align}
Throughout this work, we define the barred functions as being evaluated on the equations of motion $\phi^{+K} = \phi^{-K}$,
\begin{align} 
\bar G'_{ij} \equiv G'_{ij} \big|_{\phi^{+K} = \phi^{-K}}.
\end{align}
Furthermore, we have defined the derivatives of $G_{ij}$ by
\begin{align}
G'_{ij} \equiv  \frac{\partial G }{ \partial \left( K^{i}_{\delta} K^{j\delta}\right) }.
\end{align} 

The first-order equations of motion for $S^{(1)}_{CTP}$  are
\begin{align}\label{EOM1}
&\partial_\lambda \sum_{i,j,k} \Big\{  i f_{ijk} P^{i\mu\lambda}_K K^{j\nu} \partial_\mu K^k_\nu  \nn
&+  j f_{ijk} K^{i\mu} P^{j\nu\lambda}_K \partial_\mu K^k_\nu  - k f_{ijk} K^{i\mu} \partial_\mu K^j_\nu P^{k\nu\lambda}_K   \nn
&+ \sum_{l\leq m} f'_{ijk,lm} \Big[ \left(l P^{l\gamma\lambda}_K K^m_\gamma + m K^l_\gamma P^{m\gamma\lambda}_K \right) K^{i\mu} K^{j\nu} \partial_\mu K^k_{\nu}  \nn
&-   k  \partial_\mu \left( K^{l}_{\gamma} K^{m\gamma} \right) K^{i\mu} K^{j}_\nu P^{k\nu\lambda}_K   \Big]  \Big\} = 0, 
\end{align}
where 
\begin{align}
f_{ijk,lm} \equiv \frac{ \partial f_{ijk} }{ \partial K^{l}_{\delta} K^{m\delta} }. 
\end{align}
The calculation of $T_{(1)}^{\mu\nu}$ goes through as it did for $T_{(0)}^{\mu\nu}$, resulting in a non-symmetric tensor $T^{\mu\nu}$ on the left-hand-side of \eqref{embalance}, 
\begin{align}\label{Tderiv}
T^{\mu\nu} =&~ \varepsilon u^\mu u^\nu+p\Delta^{\mu\nu} - \eta_1 u^\mu u^\lambda\partial_\lambda u^\nu \nn
&+ \left( \chi_1 \eta^{\mu\nu}\! + \chi_2 u^\mu u^\nu \right) \partial_\lambda u^\lambda + \beta u^\mu \partial^\nu b,
\end{align}
where the coefficient functions are given by
\begin{align}
&\eta_1 = \frac{b^3}{3} \sum_{i,j,k} \left(j-k\right) \bar f_{ijk}, \label{eta} \\
&\chi_1 = \chi_2 + b \beta , \label{chi1} \\
&\chi_2 =  \frac{b^3}{3}    \sum_{i,j,k}   \sum_{l\leq m}  \left[ \left(j-k\right) \bar f_{ijk} -  C_{ijk,lm} \right] ,\label{chi2} \\
&\beta =  \frac{b^2}{3} \sum_{i,j,k} i \bar f_{ijk} , \label{beta}
\end{align}
with 
\begin{align}
C_{ijk,lm}\equiv  b^2  \bar f'_{ijk,lm}  \left( l +  m  - 2 k \right) . 
\end{align}
The contribution to the non-conserving $R^{\nu}$ from the first-order action is  
\begin{align}\label{force1}
\! R_{(1)}^{\nu} =  \eta_1 u^\alpha \partial_\alpha u_\lambda \partial^\nu u^\lambda + \frac{\chi_1}{b} \partial_\lambda u^\lambda \partial^\nu b - \frac{\beta}{b} \partial_\lambda b \partial^\nu u^\lambda  .
\end{align}

\subsection{Stress-energy tensor, the Navier-Stokes equations and the bulk viscosity}
The most important remaining question is how the energy-momentum balance equation \eqref{embalance} relates to viscous phenomenological hydrodynamics, which can be obtained from the symmetric stress-energy tensor
\begin{align}
T^{\mu\nu}_{ph} = T^{\mu\nu}_{(0)ph} + T^{\mu\nu}_{(1)ph}.
\end{align}
The form of $T^{\mu\nu}_{(0)ph}$ equals that of $T^{\mu\nu}_{(0)}$ in Eq. \eqref{TnonCTP} and 
\begin{align}\label{Tpheno}
T^{\mu\nu}_{(1)ph} &= - \eta \sigma^{\mu\nu} - \zeta \Delta^{\mu\nu} \partial_\lambda u^\lambda + \left(q^\mu u^\nu + q^\nu u^\mu \right).
\end{align}
The tensor $\sigma^{\mu\nu}$ is the transverse traceless symmetric tensor
\begin{align}
\sigma^{\mu\nu} \equiv \Delta^{\mu\alpha} \Delta^{\nu\beta} \left( \partial_\alpha u_\beta + \partial_\beta u_\alpha - \frac{2}{3} \eta_{\alpha\beta} \partial_\lambda u^\lambda \right).
\end{align}
The tensorial structures in Eqs. \eqref{TnonCTP} and \eqref{Tpheno} directly follow from stress-energy tensor $T^{\mu\nu}_{ph}$, which can be written as
\begin{align}\label{Tframe}
T^{\mu\nu}_{ph} = \CE u^\mu u^\nu + \CP \Delta^{\mu\nu} + \left( u^\mu q^\nu + u^\nu \tilde q^\mu   \right) + t^{\mu\nu}, 
\end{align}
with $q^\mu$ and $\tilde q^\mu$ transverse, and $t^{\mu\nu}$ transverse, symmetric and traceless. Because of the symmetry of the phenomenological hydrodynamic stress-energy tensor, it is clear that $\tilde q^\mu = q^\mu$. The two scalars $\CE$ and $\CP$, the vector $q^\mu$ and the tensor $t^{\mu\nu}$ are then constructed in terms of the gradient expansion in temperature, chemical potential and velocity fields: $T(x)$, $\mu(x)$ and $u^\mu(x)$ (see e.g. \cite{Kovtun:2012rj,Bhattacharya:2011tra}). In our discussion of neutral fluids, $\mu(x) = 0$.

Despite the fact that the tensor $T^{\mu\nu}$ we derived in \eqref{Tderiv} is not conserved, we can write it in the form of \eqref{Tframe}. It is important to note that $T^{\mu\nu}_{(1)}$ is {\it not} symmetric, thus $\tilde q^\mu \neq q^\mu$. At this point, the fact that $T^{\mu\nu}$ is not symmetric means that we cannot interpret it as a stress-energy tensor, in the absence of the Belinfante-Rosenfeld procedure \cite{Belinfante1940449,Rosenfeld}. However, we will see below that within an approximate scheme, a simple symmetrisation of $T^{\mu\nu}$ can lead to a hydrodynamic stress-energy tensor, which reproduces exactly the same physical equations as the ones we have derived from the energy-momentum balance equation \eqref{embalance}.

The tensor structure of \eqref{Tframe} allows us to identify the coefficient functions of \eqref{Tderiv} as
\begin{align}
\CE &= u_\mu u_\nu T^{\mu\nu} = \varepsilon  , \label{Econstr} \\ 
\CP &=  \Delta_{\mu\nu} T^{\mu\nu} / 3 = p + \chi_1 \partial_\lambda u^\lambda, \label{Pconstr} \\
q_\mu &= - \Delta_{\mu\beta} u_\alpha T^{\alpha\beta} \nn
&=  \beta \partial_\mu b - b \beta  u_\mu \partial_\lambda u^\lambda - \eta_1 u^\lambda \partial_\lambda u_\mu , \\
\tilde q_\mu &= - \Delta_{\mu\alpha} u_\beta T^{\alpha\beta}\! = 0 , \\
 t_{\mu\nu}  &= \frac{1}{2}  \left[ \Delta_{\mu\alpha} \Delta_{\nu\beta} +  \Delta_{\mu\beta} \Delta_{\nu\alpha} - \frac{2}{3} \Delta_{\mu\nu} \Delta_{\alpha\beta} \right]  T^{\alpha\beta} \nn
&= 0 \label{etaconstr}.
\end{align}

Phenomenological stress-energy tensor is by construction conserved. Its conservation equations, 
\begin{align}
&\partial_\mu T^{\mu 0}_{ph} = 0, \\
&\partial_\mu T^{\mu i}_{ph} = 0,
\end{align}
give the continuity equation and the Navier-Stokes equation, respectively. They can be reduced to their standard compressible form by using the non-relativistic scaling \cite{Bhattacharyya:2008kq}: $t\to t/\epsilon_{nr}^2$, $x\to x/\epsilon_{nr}$, $v^i \to \epsilon_{nr} v^i$ and $p\to\epsilon_{nr}^2 p$,
\begin{align}
\!\!\!&\partial_0 \rho + \partial_i \left(\rho v^i\right) = 0, \label{NSeq1} \\
\!\!\!&\rho \left( \partial_0 + v^j \partial_j \right) v^i \!= - \partial^i p + \eta \partial^2 v^i \!+ \! \left(\zeta + \eta /3 \right) \partial^i \partial_j v^j  \! ,  \label{NSeq2}
\end{align}
where $v^i$ is the velocity field, $\rho = \varepsilon + p$ with $\varepsilon$ the energy density and $p$ pressure, and $\partial^2 = \partial^j \partial_j$. In the scaling relations above, $\epsilon_{nr}$ is an infinitesimally small parameter that is taken to zero in order to find the dominant non-relativistic terms.

To show how \eqref{NSeq1} and \eqref{NSeq2} arise in our construction, we first note that the effective Goldstone action \eqref{LagCTP1}, \eqref{LagCTP2} for $\phi^\pm$ fields describes an {\it out-of-equilibrium} theory in which the gradient expansion is organised by counting derivatives of currents $K^{i\mu}$ at some IR hydrodynamic scale $\Lambda_h$. To understand the {\it near-equilibrium} limit, we study the energy-momentum balance equation \eqref{embalance} by introducing a near-equilibrium parameter $\ell$, so that 
\begin{align}
\phi^I(x) \!= b_0^{1/3} \left(x^I + \ell \pi^I(x) \right).
\end{align}
Expanding around a constant equilibrium current 
\begin{align}
K^\mu_0 = \left(b_0,0,0,0\right), 
\end{align}
it follows that
\begin{align}
&b = b_0 + \ell \Delta b + \ldots , \label{bNearEq}\\
&u^\mu = u_0^\mu + \ell v^\mu + \ldots   \label{uNearEq},
\end{align}
with 
\begin{align}
u^\mu_0 = \left(1,0,0,0\right), && v^\mu = \left(0,v^i\right). \label{Defu0andv}
\end{align}
In terms of the fluctuation fields $\pi^i$, we find that 
\begin{align}
&\Delta b =b_0 \partial_i \pi^i, \label{bvpirel1} \\
&v^i = -\partial_0 \pi^i . \label{bvpirel2}
\end{align}
The conservation equation \eqref{Kconser} then implies the order-$\ell$ relation
\begin{align}\label{Kconser1}
b_0 \partial_i v^i = - \partial_0 \Delta b.
\end{align}
At the leading order in $\ell$, the force of the environment acting on the fluid that is encoded in the non-conserving $R^\nu_{(1)}$ vanishes. The first-order $T^{\mu\nu}_{(1)}$ is thus approximately conserved near equilibrium and can be treated as the viscous contribution to the total fluid's conserved two-tensor $T^{\mu\nu}$. 

Since first-order contributions are suppressed in the double expansion by $\ell$ as well as a derivative acting on $v^i$, we expand the zeroth-order energy-momentum balance equation to order $\ell^2$. The contribution from $R^{\nu}_{(0)}$ remains non-vanishing, but it can be absorbed into the small $\CO(\ell)$-suppressed shifts of the fluid's energy and pressure,
\begin{align}
\varepsilon \to \varepsilon + \ell p_0, & &p \to p - \ell p_0,
\end{align}
where the un-shifted expressions are those of Eqs. \eqref{CTP0coeff1} and \eqref{CTP0coeff2}. Furthermore, $p_0$ is given by the expression 
\begin{align}
\!p_0 = \frac{ \left(i+j\right)}{3} \Delta b \left[ b_0 \bar G'_{ij} + \frac{1}{2} \ell \left(\bar G'_{ij}  + b_0 \partial_b \bar G'_{ij} \right)  \Delta b \right],
\end{align}
with $G'_{ij}$ evaluated at $b=b_0$ and the expression summed over $i$ and $j$.

With this re-definition of $\varepsilon$ and $p$, the tensor $T^{\mu\nu}$ in \eqref{Tderiv} becomes approximately conserved near equilibrium and mimics the expected behaviour of a {\it stress-energy tensor},
\begin{align}\label{ApproxConsOfT} 
\partial_\mu T^{\mu\nu} \approx 0.
\end{align}
A further requirement for a genuine identification of $T^{\mu\nu}$ with the hydrodynamic stress-energy tensor of the fluid described by our CTP construction, is that $T^{\mu\nu}$ needs to be symmetric. We can show that to the order of $\ell$ we are working at, a symmetrised $T^{(\mu i)}$ obeys 
\begin{align}
\partial_\mu T^{\mu i} = \partial_\mu T^{(\mu i)} = \frac{1}{2} \partial_\mu \left( T^{\mu i} + T^{i\mu} \right) + \CO(\ell^2) \approx 0.
\end{align}
The symmetrisation of $T^{\mu 0}$ does not work in the same way. However, in the non-relativistic limit, only zeroth-order, ideal hydrodynamic terms of the $T^{\mu 0}$ components contribute to the continuity equation \eqref{NSeq1}. Thus, for a non-relativistic, near-equilibirum Navier-Stokes fluid, we can identify the symmetrised version of our tensor $T^{\mu\nu}$ with the phenomenological stress-energy tensor,
\begin{align}
T^{(\mu\nu)} \approx T^{\mu\nu}_{ph}.
\end{align}  
One should be aware that beyond the aesthetic desire to exactly match the phenomenological stress-energy tensor, what is important for the physics are the dynamical equations of motion. Those follow from Eq. \eqref{embalance}, which is approximately conserved and does not require $T^{\mu\nu}$ to be symmetric. The dynamical equations derived in the near-equilibrium limit of our CTP construction are thus completely equivalent to those derived from phenomenological hydrodynamics with the use of conservation laws.

The Navier-Stokes equations \eqref{NSeq1} and \eqref{NSeq2} again follow from the near-equilibrium expansion to $\CO(\ell^2)$ at zeroth order, and $\CO(\ell)$ at first order in gradient expansion, followed by a non-relativistic scaling limit $\epsilon_{nr}\to 0$. From this expansion, or directly from \eqref{etaconstr}, we find that the shear viscosity $\eta$ vanishes while the bulk viscosity is non-zero,
\begin{align}
\eta = 0, & & \zeta = - \chi_1 |_{b=b_0}.
\end{align}
Note that the vanishing of the shear viscosity is most likely caused by the very large symmetry group of volume-preserving diffeomorphisms, under which our fluid is invariant. In fact, viscosity in \cite{Endlich:2012vt} resulted from a Lagrangian term that explicitly broke this symmetry. Furthermore, the non-dissipative construction of second-order hydrodynamics invariant under $S\text{Diff}\,(\mathbb{R}^{1,3})$ in \cite{Bhattacharya:2012zx} also resulted in a smaller number of independent transport coefficients compared to the phenomenological classification, which is based on the counting of independent tensor structures.

Because of the near-equilibrium expansion, the hydrodynamic coefficient $\zeta$ becomes an equilibrium, $b_0$-dependent constant. In terms of the four undetermined coefficient functions in the Lagrangian \eqref{LagCTP2},  
\begin{align}\label{ZetaResult}
\zeta &= - b^3_0 \left(\bar f_{333} +\bar f_{300} - \bar f_{303} + 3 \bar f_{330} \right) |_{b=b_0} - 2 b_0^5 \left(  \bar f'_{333,03} \right. \nn
&\left.+ \bar f'_{303,03} - \bar f'_{330,03} - \bar f'_{300,03}  \right)|_{b=b_0} - 4 b_0^5 \left(  \bar f'_{333,00} \right. \nn
&\left. + \bar f'_{303,00} \right) |_{b=b_0} + 4 b_0^5 \left(  \bar f'_{330,33} + \bar f'_{300,33} \right) |_{b=b_0} .
\end{align}

\subsection{The entropy current}

Lastly, let us turn our attention to the entropy current $S^\mu$, which can be associated with the system.\footnote{We thank the anonymous PRD referee for insightful comments on the topic of the entropy current.} In standard relativistic hydrodynamics with a conserved stress-energy tensor, the entropy current can be written as a sum of two terms,
\begin{align}
S^\mu = S^\mu_{can} + S^\mu_{corr}.
\end{align}
The first term, $S^\mu_{can}$, is the canonical part that must satisfy the covariant relativistic generalisation of the thermodynamical relation $\varepsilon + p = s T$ \cite{Kovtun:2012rj},
\begin{align}\label{EntCurrGeneral}
T S^\mu_{can} = p u^\mu - T^{\nu\mu} u_\nu.
\end{align} 

It is important to note that the canonical entropy current is invariant to first order in $\partial_\mu u_\nu$ and $\partial_\mu T$ under the frame re-definitions $u^\mu \to u^\mu + \delta u^\mu (\partial_\alpha u_\beta, \partial_\alpha T) $ and $ T\to T + \delta T (\partial_\alpha u_\beta, \partial_\alpha T)$. 

In the spirit of the gradient expansion of hydrodynamics, one should be allowed to add the most general series of corrections \cite{Loganayagam:2008is,Bhattacharyya:2008xc}, as terms in $S_{corr}^\mu$, that are consistent with the symmetries of the vector current $S^\mu$ and ensure the positivity of the entropy production, 
\begin{align}\label{PosEntProd}
\partial_\mu S^\mu \geq 0.
\end{align}
However, for our purposes, $S_{corr}$ is irrelevant as it has long been known that corrections to $S_{can}$ can only arise at the second order in the derivative expansion of the stress-energy tensor, which are beyond the scope of this work \cite{DeGroot:1980dk,Loganayagam:2008is,Bhattacharyya:2008xc,Romatschke:2009kr,Bhattacharyya:2012nq}.\footnote{In more general hydrodynamic frames, not considered in this work, first-order corrections to \eqref{EntCurrGeneral} may appear in the expansion of the entropy current. For some recent discussions on those topics and various related hydrodynamic extensions, see \cite{Becattini:2014yxa,Hayata:2015lga,Van:2011yn} and references therein.} 

Although associating an entropy current is difficult in our situation when the full non-linear theory is considered, the linearised stress-energy tensor was shown to be approximately conserved near equilibrium, cf. \eqref{ApproxConsOfT}. If we restrict ourselves to work only within that regime,  then Eq. \eqref{EntCurrGeneral} implies that
\begin{align}\label{EntCurr}
S^\mu = s u^\mu + \frac{ q^\mu }{ T } =\left( \frac{\varepsilon + p }{T }\right) u^\mu  + \frac{q^\mu }{ T} .
\end{align} 

By using the standard thermodynamical expressions $\varepsilon + p = s T$, $T = \partial \varepsilon / \partial s$ and $s = \partial p / \partial T$, along with Eqs. \eqref{CTP0coeff1} and \eqref{CTP0coeff2}, we find the temperature and the entropy density to be 
\begin{align}
&T = \exp\left\{ - \ln C + \CI  \right\} , \label{TDis} \\
&s = \exp\left\{\ln\left[ - C b\partial_b F - \frac{1}{3} C b^2 \CG  \right] - \CI  \right\},  \label{sDis} 
\end{align} 
where $C$ is an integration constant and we have defined
\begin{align}
&\CI \equiv \int db \left[\frac{\partial^2_b F - \frac{2}{3} \CG - \frac{1}{3} b \partial_b \CG }{\partial_b F - \frac{1}{3} b \CG}  \right],\\
&\CG \equiv \sum_{i\leq j } \bar G'_{ij} \left(i+j\right).
\end{align}
In the absence of dissipation, when $G(K^i_\gamma K^{j\gamma}) = 0$, the expressions \eqref{TDis} and \eqref{sDis} reduce to the previously derived non-dissipative results, $T  = - \partial_b F$ and $s=b$, after we set $C=-1$.

Let us proceed by considering the implications of imposing Eq. \eqref{PosEntProd}, i.e. the positive entropy production condition. A conserved hydrodynamical stress-energy tensor \eqref{Tframe} with $\tilde q^\mu = q^\mu \neq 0$ and $\eta = 0$ leads to an equation of motion, $u_\nu \partial_\mu T^{\mu\nu}_{ph} = 0$, which can be written out in terms of the entropy density,
\begin{align}\label{EoMEnt}
\partial_\mu \left(s u^\mu \right) = \frac{1}{T} \zeta \left(\partial_\mu u^\mu\right)^2 - \frac{1}{T} \partial_\mu q^\mu + \frac{1}{T} u_\nu u^\mu \partial_\mu q^\nu.
\end{align}

Eq. \eqref{EoMEnt} can now be used to eliminate the entropy density term from the divergence of \eqref{EntCurr}. The resulting inequality can be written out purely in terms of the constituents of the CTP Lagrangian. Although we will not consider this expression here in detail, what is important is that the action itself does not guarantee the entropy production to be positive and additional constraints must be imposed on the form of $G$ and $f_{ijk}$ to ensure a physically sensible effective theory. 

The same conclusion can be drawn by studying the sign of the bulk viscosity $\zeta$ in Eq. \eqref{ZetaResult}, which would have to be non-negative in a physical fluid.\footnote{Microscopically, the positivity of the two viscosities ($\eta$ and $\zeta$) is ensured by the structure of the real-time two-point Green's functions, which appear in the Kubo formulae. For a more detailed discussion, see e.g. \cite{Kovtun:2014hpa}.} Again, imposing $\zeta \geq 0$ results in constraints on the form of the $f_{ijk}$ functions.

The fact that we need to impose additional restrictions on the form of the effective action in order for the system to produce positive entropy and have non-negative energy density, temperature, bulk viscosity, etc., is not an unexpected feature of our construction. It results from the fact that the ``phenomenological" effective action with the Lagrangian \eqref{LagCTP1} and \eqref{LagCTP2} was not derived from a unitary, microscopic quantum field theory. Had we done this, the structure of the Schwinger-Keldysh propagators would ensure that such problems would not be present in the effective infrared theory.\footnote{See \cite{Grozdanov:2015nea} for a recent derivation of an effective CTP action for the Noether current from quantum electrodynamics and discussions therein on how the structure of the Schwinger-Keldysh propagators ensures a consistent IR effective theory. The precise mechanism for how this microscopic structure should be implemented in effective theories of the type studied in this work remains to be understood. Recently, some of these issues were also discussed in \cite{Haehl:2015pja}.}

Finally, as a simpler example, consider a very special family of Lagrangians (or fluid flows) in which $G(K^i_\gamma K^{j\gamma}) = 0$, so that the zeroth-order part of the entropy current remains equals to $K^\mu$, which is conserved by construction. The positivity of the divergence of \eqref{EntCurr} would then require us to only impose 
\begin{align}\label{EntProdCTP}
\partial_\mu \left( \frac{q^\mu }{ T } \right) \geq 0. 
\end{align}
It is now easy to show that at the leading order in $\ell$ and in the non-relativistic limit, that positive entropy production condition \eqref{EntProdCTP} demands that 
\begin{align}\label{EntProdFirstOrder}
\beta(b_0) \partial^i \partial_i \Delta b \geq 0.
\end{align} 
Since $\beta(b_0) = b_0^2 \sum_{i,j,k} i \bar f_{ijk} (b_0) / 3$, Eq. \eqref{EntProdFirstOrder} explicitly shows that we need to supply additional constraints on $f_{ijk}$ to ensure positive entropy production. 

It is interesting to note that the expression \eqref{EntProdFirstOrder} is consistent with the following fact pertaining to incompressible fluids, which are characterised by the condition 
\begin{align}\label{NRIncompress}
\partial_i v^i = 0. 
\end{align}
According to the definitions \eqref{uNearEq} and \eqref{Defu0andv}, the incompressibility condition \eqref{NRIncompress} implies the relativistic relation, $\partial_\mu u^\mu = 0$, to first order in $\ell$. The conservation of $K^\mu$, cf. Eq. \eqref{Kconser}, then implies that $b$ must be a space-time independent constant. Given the definition \eqref{bNearEq} of $b$ to order $\ell$, the fact that $b$ must be constant means that we may absorb a constant value $\Delta b$ into $b_0$, and set $\Delta b = 0$. Finally, Eq. \eqref{EntProdFirstOrder} shows that incompressibility implies conservation of entropy. These findings are therefore consistent with the fact that an incompressible non-relativistic fluid with $\eta = 0$ behaves as an ideal fluid without any entropy production. In such cases, the presence of the bulk viscosity $\zeta$ alone cannot influence the solutions of the Navier-Stokes equation \eqref{NSeq2}.

\section{ Conclusion } 

In this work, we showed how phenomenological relativistic hydrodynamics with dissipation can be constructed using a classical CTP effective action. We were able to derive closed-form equations describing the fluid from an action principle, containing dissipative effects triggered by the presence of the non-zero bulk viscosity. 

Of central importance were terms collected into $S_2$, which coupled fields living on the two time axes and reflected quantum and classical interactions between the open (sub)-system and the integrated-out, UV degrees of freedom of the environment. Such terms were argued to generically arise in an effective CTP field theory. Dissipation thus manifested itself in the energy loss of the low-energy degrees of freedom to the UV microscopic degrees of freedom. We note that this physical interpretation is in accordance with the usual phenomenological view of dissipation. However, in that approach one is able to maintain all conservation laws. Despite the immense historical success of such a phenomenological approach, the fact that energy should {\it not} be conserved in a theory describing only the relevant hydrodynamic modes, is a natural result of interactions between all degrees of freedom before a choice is made to eliminate some of them from our description of the system. Our future plan is to explore how this effective theory point of view could be related in a more precise and quantitative manner to the phenomenological assumption of conservation laws.   

Despite the lack of energy conservation in our {\it open} effective field theory, the two-tensor $T^{\mu\nu}$ we derived was shown to be conserved in the near-equilibrium regime, thus approaching the behaviour of the phenomenological stress-energy tensor in that limit. This enabled us to identify the bulk viscosity of the family of fluids that could be described by the action we constructed. The shear viscosity, however, vanished in this setup, which is most likely the result of a large amount of symmetry, namely the volume preserving diffeomorphisms that were used to construct the effective action. We defer a further study of this problem, i.e. the identification of the correct symmetries for a description of dissipative fluids, as well as the classification of different fluids described by the presented formalism to the future work.  

The systematic treatment of dissipation we presented in this paper can be applied to any effective theory derived from a quantum field theoretical model. The thermodynamical considerations of the usual phenomenological approach are thus completely replaced by the assumption about the applicability of the Wick's theorem and the gradient expansion. We saw that the positivity of the divergence of the dissipative entropy current, as defined in this work, was not automatically ensured and additional restrictions would need to be imposed on the form of the effective action. A more detailed future investigation into how the microscopic CTP structures constrain infrared effective theories will be required to fully resolve this problem.

\section*{Acknowledgements}

The authors would like to thank Sergei Dubovsky, Nikolaos Kaplis, David Kralji\'{c}, Alberto Nicolis and Andrei Starinets for useful discussions and comments on the manuscript. S.~G. is supported by the Graduate Scholarship of St. John's College, Oxford.

\bibliography{HydroCTPRefs}

\end{document}